# Unlocking Insights into Business Trajectories with Transformer-based Spatio-temporal Data Analysis

**Muhammad Arslan[1], Christophe Cruz[2]**

1. *Laboratoire Interdisciplinaire Carnot de Bourgogne (ICB),*
   *9 Av. Alain Savary, 21000 Dijon, France*
   *muhammad.arslan@u-bourgogne.fr*

2. *Laboratoire Interdisciplinaire Carnot de Bourgogne (ICB),*
   *9 Av. Alain Savary, 21000 Dijon, France*
   *christophe.cruz@u-bourgogne.fr*

*ABSTRACT. The world of business is constantly evolving and staying ahead of the curve requires a deep understanding of market trends and performance. With the increasing availability of text data, it has become possible to gain insights into the health and trajectory of businesses and help decision-makers to make informed decisions. This work is based on the application of the Transformer-based spatio-temporal data analysis for analyzing vast amounts of business-related text data. The proposed method processes news articles that were retrieved from multiple sources from 2017 to 2021 inclusive. It categorizes them into different business categories (e.g. manufacturing, healthcare, construction, etc.) based on the company's interest. A company's taxonomy is used to understand their domains of interest. It then identifies the main topics in each article, which give us information about the events (e.g. launch of a company, regulatory changes, etc.) and their spatio-temporal evolution over time. The insights generated from this analysis can be used by businesses to make informed decisions, by investors to identify opportunities and assess risk, and by policymakers to understand the impact of regulations and economic policies on the business community.*

# 1. Introduction

Transformer-based spatio-temporal data analysis is a cutting-edge approach that leverages the power of machine learning and natural language processing to analyze vast amounts of business-related text data (Fan et al. 2022). This approach allows us to not only analyze the performance of businesses over time but also understand how trends and performance vary across geographic territories. By combining data analysis with the latest advancements in natural language processing, we can gain a comprehensive view of business trends (Brașoveanu and Andonie 2020). It offers a powerful tool for unlocking insights into business trajectories, providing valuable information for businesses, investors, and policymakers.

To perform business data analysis, we need to develop a news data analyzer (Alawadh et al. 2023). A news data analyzer refers to a system that processes and analyzes news articles to extract relevant information and insights (Lau et al. 2021). The goal of a news data analyzer is typically to gain a better understanding of the events and trends in the news and to make sense of the large amounts of news data that are generated daily. A news data analyzer can use various techniques and algorithms (Ashtiani and Raahmei 2023), such as Natural Language Processing (NLP), Machine Learning (ML), and Data Mining, to process news articles and extract valuable information such as topics, etc. The analyzer can then present this information in a meaningful and useful way, such as through visualizations. The specific features and capabilities of a news data analyzer can vary depending on the use case and requirements, but the overall goal is to help users make sense of the news data and gain valuable insights and information.

In our case, a news data analyzer is a system that combines transformer-based text classification (Cunha et al. 2022) and dynamic topic models (Grootendorst 2022) and classifies news articles into different business categories. Business categories refer to different types of businesses (e.g., healthcare, transportation, agriculture, construction, and real estate) that exist in the market. For this study, the domains were shortlisted based on the company's interest. After categorizing the business news, it identifies the main topics in each text. In the context of business, many different topics can be covered in news articles. For instance, company news and performance, which deals with updates on the performance and financial results of individual companies, as well as news on mergers and acquisitions, partnerships, and other business developments. The insights generated by understanding the spatio-temporal evolution of these topics will help to make informed decisions.

This paper is organized as follows. Section 2 reviews the background of transformer-based text classification and dynamic topic models. Section 3 introduces the proposed method. The discussion is mentioned in Section 4 and the conclusion is described in Section 5.

## 2. Background

Text classification is a task in natural language processing (NLP) where the goal is to assign predefined labels to a given text based on its content (Kowsari et al. 2019). The traditional approach to text classification was based on hand-engineered features, such as the presence of certain keywords, and simple machine learning algorithms like Naive Bayes (Webb et al. 2010). With the advent of deep learning, the focus shifted towards end-to-end models that could automatically extract useful features from the raw text. One of the most successful such models is the Transformer, introduced in 2017 (Rothman 2021).

Transformers are a type of neural network architecture that use self-attention mechanisms to process input sequences in a parallel manner (Rothman 2021). In the context of NLP, they have proven to be highly effective at capturing long-range dependencies and semantic relationships between words in a sentence, which is essential for accurate text classification. Transformers have been applied to various NLP tasks, including text classification (Bhardwaj et al. 2021). In text classification, the input text can be assigned one or multiple labels (Liu et al. 2021). If a text can be assigned multiple labels, the task is known as multi-label text classification. Transformers can be applied to both single-label and multi-label text classification tasks. In the case of multi-label text classification, the output of the model is a binary vector, with each dimension corresponding to a label (Liu et al. 2021). The value of each dimension is either 0 or 1, indicating whether the text belongs to that label. To train a Transformer for multi-label text classification, a binary cross-entropy loss is used (Hande et al. 2021). The model is trained to predict the probability of each label, and the binary cross-entropy loss measures the difference between the true label and the predicted probability.

Dynamic topic models (DTMs) are a type of generative probabilistic model that can be used to analyze and track the evolution of topics in large collections of text documents over time (Grootendorst 2022). In traditional topic models, such as Latent Dirichlet Allocation (LDA), topics are treated as fixed latent variables that generate the words in the documents (Blei et al. 2003). However, in dynamic topic models (e.g., BERTopic), topics are assumed to evolve over time, reflecting the changing interests and themes of the underlying text corpus. DTMs typically represent each document as a mixture of topics, and each topic is modeled as a probability distribution over words (Grootendorst 2022). The model then assigns topics to the words in each document, and the topic distributions for each document are used to infer the topic evolution over time. DTMs are a flexible and powerful tool for analyzing text data over time and can be used to identify trends, track the spread of ideas and information, and understand how topics change and evolve (Yao and Wang 2022). They have been applied to a wide range of text-based applications (Savin and Konop 2022), including social media analysis, news analysis, and scientific literature analysis (Sestino and De Mauro 2022).

In the existing literature, the intersection of transformer-based models and dynamic topic models (DTMs) has not been thoroughly investigated in the domain of business analysis. This research aims to fill this gap by exploring the benefits of combining these two approaches for business analysis. The objective of this study is to show the efficacy of combining transformer-based models and DTMs for multi-label text classification and topic extraction, and to demonstrate the potential applications of this combination in the realm of business analysis.

## 3. Proposed Method

The proposed news data analyzer consists of four key stages:

1) Text preprocessing involves cleaning, transforming, and preparing text data for spatio-temporal analysis.
2) Text classification involves categorizing text data into predefined categories based on their content and characteristics.
3) Dynamic topic modeling involves identifying and tracking the evolution of topics over time within a text corpus.
4) Data visualization is the final stage, where the results of the previous stages are presented in an intuitive and graphical manner to facilitate understanding and interpretation of the business trajectories and their evolution over time and geographic regions.

### *3.1. Text preprocessing*

It is a crucial step in the NLP pipeline that involves cleaning and transforming raw data into a format suitable for analysis. The goal of preprocessing is to improve the quality and reliability of the data and to facilitate the analysis by reducing the amount of noise and irrelevant information. The choice of preprocessing techniques depends on the specific dataset and the problem being solved, and it can have a significant impact on the performance of the final model. For our case, a dataset of 26,509 news articles is used. This dataset is manually labelled using 1,800 different categories based on the company's taxonomy. An example is shown in Fig. 1. The text is classified with labels. These labels correspond to the labels of the taxonomy. The news articles were preprocessed, and a corpus of around 8,128,981 words is obtained after preprocessing the news articles. The system implements several preprocessing functions for text data in the English language. It starts by importing the required libraries including NLTK, RE, and String. It first removes punctuation and digits from the text using regular expressions. Then, it converts the text to lowercase, tokenizes the text into words, removes stop words and duplicates, and removes small words. The final step is to join the preprocessed words into a single text to feed the classification model.

```
company building ventelle transport area
specialize corbehem consider logistics employee
move want brebières second built site base
['112', '4716', '25276']
```

| LabelID | LabelDefinition |
|---|---|
| 25276 | transport and logistics |
| 112 | construction of premises |
| 4716 | moving house |

FIGURE 1. *Text labelled using the taxonomy of the company.*

*3.2. Multi-label text classification*

As each text is labelled with multiple labels (see Fig. 1) in the dataset, the case of multi-label classification is followed. After preprocessing the text, a transformer-based multi-label text classification model is executed. The model uses a multi-layer neural network architecture and is trained using the binary cross-entropy loss function and the Adam optimizer. The code imports several libraries, including Numpy and Pandas for data processing and manipulation, and Keras, a high-level deep learning library, for building and training the model. The model starts by loading the preprocessed text data using Numpy. This data is then tokenized using the Tokenizer class from Keras. Later, the tokenized text data is converted into numerical sequences. The sequences are then padded to make all sequences the same length. This is done so that all the input data has the same shape, which is required for training deep learning models. The data is then split into training and validation sets using Numpy slicing. The split index is calculated as 80% of the total number of sequences.

Next, the deep learning model is defined using the Keras functional API. The model consists of several layers, including an input layer, an embedding layer, a global max pooling layer, a dense layer with a Rectified Linear Unit (ReLU) activation function, and a dropout layer. The dropout layer helps to prevent overfitting of the model by randomly dropping out a fraction of the neurons during each training step. The final layer is a dense layer with a sigmoid activation function, which produces binary output values. The model is then compiled, where the loss function, optimizer, and evaluation metrics are specified. The loss function used is binary cross-entropy, which is suitable for binary classification problems. The optimizer used is the Adam optimizer with a learning rate of 0.001. The evaluation metric used is accuracy and precision. Finally, the model is trained using the *fit* method. The training data is used along with several other parameters such as batch size and number of epochs. The validation data is also used to evaluate the performance of the model.

*3.3. Dynamic topic modeling*

To use BERTopic, a transformer-based dynamic topic modeling approach, we first convert the preprocessed news text into numerical data using the "sentence-transformers" python package. This package converts text data into 512-dimensional vectors, providing document-level embeddings. We set the language to English since our corpus is in English. During training of the BERTopic model, topic probabilities are calculated. Once training is completed, the model outputs 66 business topics, each represented by the 10 most probable words (see Fig. 2).

To visualize the evolution of topics in news articles over time, we use the "*topic_model.topics_over_time*" function, which requires the preprocessed news text corpus (docs), acquired topics (topics), timestamps for each article, global tuning for averaging topic representation with global topic representation, evolution tuning for averaging topic representation with the previous time's representation, and the number of bins for timestamps (nr_bins). The input values result in a plot (Fig. 3) showing the evolution of business topics over the past 5 years.

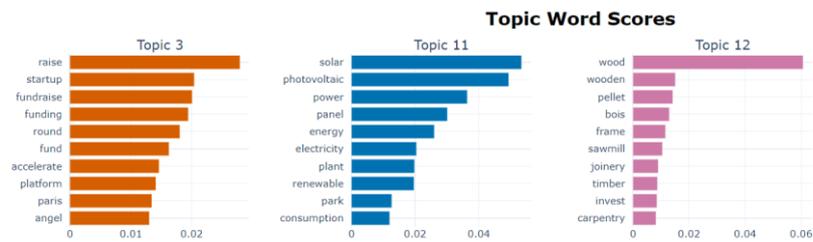

FIGURE 2. *Different words representing different topics.*

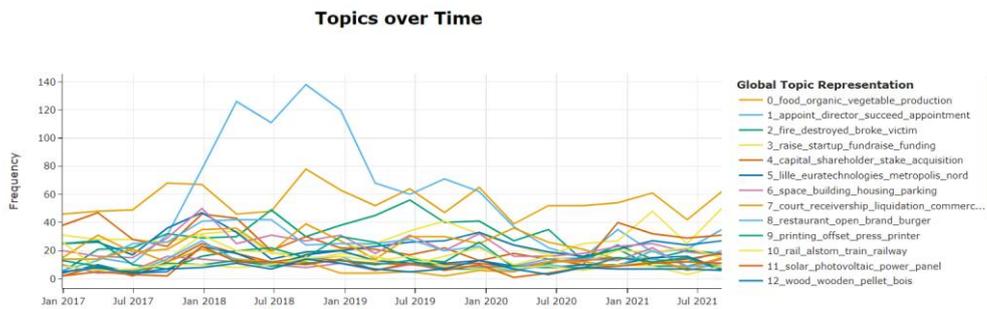

FIGURE 3. *The evolution of different topics over time.*

### 3.4. Visualizations of business trajectories

The process of constructing business trajectory visualizations involved first performing the multi-label classification of news text. Once the text is categorized and labelled, it is important to extract the topics for understanding different types of business events. For instance, the launch of the solar energy industry represents a significant development in the field of renewable energy. This business event signifies the introduction of a new player in the market, offering innovative solutions for harnessing the power of the sun. The event is typically categorized under topic 11 of renewable energy and can be identified using keywords such as "solar," "power," "energy," "park," and others that are relevant to the industry (see Fig. 2).

It is crucial to analyze the volume (i.e., number) of news articles related to the launch of the solar energy industry. This information can provide valuable insights into the origin and evolution of the business event. By examining the number of articles published over different time periods, we can determine the level of public interest and media coverage surrounding the event. Additionally, tracking the geographic regions that the articles pertain to can give us an idea of the location and spread of the industry's development. This information can help us to identify the key players and driving forces behind the growth of the solar energy industry, providing valuable insights into the market and its potential for future growth. This information regarding the volume of news articles (see Fig. 4) was then utilized to create a business trajectory visualization (see Fig. 5) to show how the business topic evolved over time across different geographic regions for a better understanding of the overall business event.

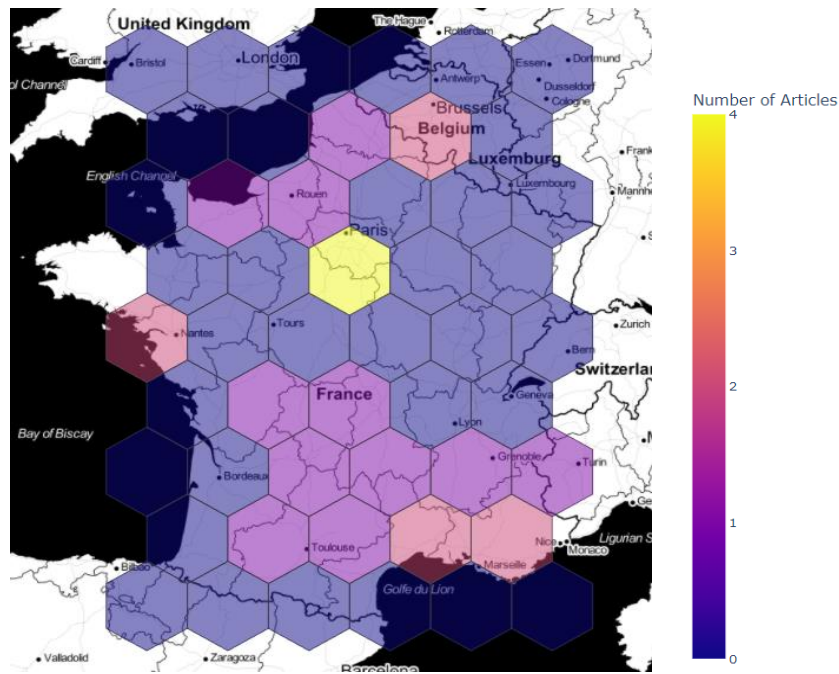

*FIGURE 4. Number of articles published on a topic.*

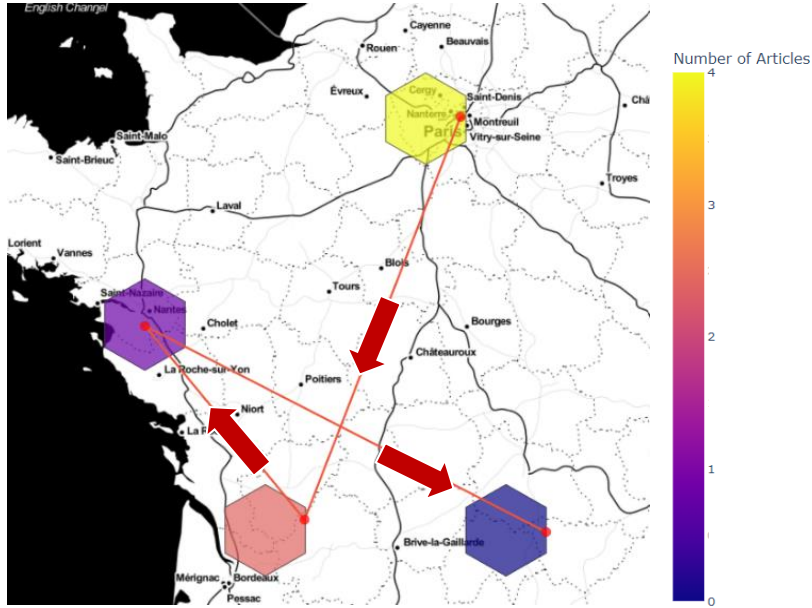

*FIGURE 5. Business trajectory across geographic regions*

**4. Discussion**

When using Transformers for multi-label text classification, there are some important considerations observed during this study. Most importantly, it is about class imbalance and label correlations. It is raised because of a huge number of labels (1800 in our case) and a small number of text examples of each label in a dataset. Class imbalance refers to the situation where some labels are much more frequent than others and can lead to a model that is biased towards the more frequent labels. To address the class imbalance, various techniques can be used, such as oversampling the minority class or using weighted loss functions. Label correlations refer to the situation where the presence of one label is often predictive of the presence of another label. This can be addressed by using techniques such as label powerset or label embedding, which represent the labels as a combination or embedding of multiple binary dimensions, respectively. However, these aspects concerning class imbalance and label correlations are not dealt with in this study.

Apart from that, visualizing business trajectories over time and geographic regions can provide several benefits (Arslan and Cruz 2023), which are:

1) *Understanding trends*: Visualizing the trajectory of a business over time helps to identify trends and patterns in its development, which can be useful for decision-making and planning.

2) *Spotting opportunities and challenges*: By seeing the business trajectory over time and space, it becomes easier to spot opportunities and challenges that the business may face. This information can be used to make informed decisions.

3) *Improved communication*: Visualizing business trajectories can help improve communication and collaboration among stakeholders, as it provides a clear and concise way to present information.

4) *Increased awareness*: By creating visual representations of business trajectories, it can increase awareness and understanding of the business among stakeholders.

5) *Enhanced analysis*: Visualizing the business trajectory over time and space allows for a more in-depth analysis of the business, as it provides a comprehensive view of its development and performance.

## 5. Conclusion

The application of Transformer-based spatio-temporal data analysis to business-related text data has proven to be an effective way to gain valuable insights into the health and trajectory of businesses. This method processes news articles and categorizes them into different business categories based on a company's interest. Later, the topics are extracted to understand different types of businesses over time. Visualizing hot and cold topics (referring to the current popularity and level of interest in a particular category) as business trajectories can greatly enhance and streamline business processes. By keeping track of these hot and cold topics, businesses can stay informed about the current state of the market and make informed decisions that align with current trends and consumer interests. This not only helps to identify potential opportunities for growth and innovation but also helps to mitigate any potential risks or challenges that may arise. Additionally, having a clear understanding of the current trends and topics can also inform and shape the direction of a company's marketing and communication strategies, allowing them to better connect with their target audience and stay relevant in the market.

## Acknowledgements

The authors thank the French company FirstECO for providing the taxonomy and the French government for the plan France Relance funding.